\documentclass[a4paper,11pt]{article}
\usepackage{graphicx}
\usepackage{epstopdf}
\usepackage{color}
\begin{document}
 
\centerline{\Large\bf Dissecting the core of the Tarantula Nebula with VLT-MUSE}

\vspace{0.5cm}

\noindent Paul A Crowther$^{1}$ \\
Norberto Castro$^{2}$ \\
Christopher J. Evans$^{3}$ \\
Jorick S Vink$^{4}$ \\
Jorge Melnick$^{5}$ \\
Fernando Selman$^{5}$ 

\vspace{0.5cm}

\noindent $^{1}$Department of Physics \& Astronomy, University of 
Sheffield, Sheffield, United Kingdom\\
$^{2}$Department of Astronomy, University of Michigan, Ann 
Arbor, USA\\
$^{3}$UK Astronomy Technology Centre, Royal Observatory, 
Edinburgh, United Kingdom\\
$^{4}$Armagh Observatory, Armagh, Northern Ireland, United Kingdom\\
$^{5}$European Southern Observatory, Santiago, Chile
\vspace{0.5cm}

{\bf We provide an overview of Science Verification MUSE observations
  of NGC~2070, the central region of the Tarantula Nebula in the Large
  Magellanic Cloud. Integral-field spectroscopy of the central
  2$'\times2'$ region provides the first complete 
    spectroscopic census of its massive star content, nebular
  conditions and kinematics. The star-formation surface density of
  NGC~2070 is reminiscent of the intense star-forming knots of
  high-redshift galaxies, with nebular conditions similar to
  low-redshift Green Pea galaxies, some of which are Lyman continuum
  leakers. Uniquely, MUSE permits the star-formation history of
  NGC~2070 to be studied from both spatially-resolved and
  integrated-light spectroscopy.}

\vspace*{0.5cm}

\noindent {\bf Tarantula nebula}

\vspace*{0.5cm}

The Tarantula nebula (30 Doradus) in the Large Magellanic Cloud (LMC)
is intrinsically the brightest star-forming region in the Local Group
and has been the subject of numerous studies across the
electromagnetic spectrum.  Its (half-solar) metallicity and high
star-formation intensity is more typical of knots in high-redshift,
star-forming galaxies than local systems, owing to a very rich stellar
content (Doran et al. 2013). Indeed, 30 Doradus has nebular conditions
that are reminiscent of `Green Peas', local extreme emission-line
galaxies that are analogues of high-redshift, intensely star-forming
galaxies, some of which have been confirmed as Lyman-continuum leakers 
(e.g. Micheva et al. 2017).

The Tarantula nebula is host to hundreds of massive stars which power
the strong H$\alpha$ nebular emission, comprising main sequence OB
stars, evolved blue supergiants, red supergiants, Luminous Blue
Variables and Wolf-Rayet (WR) stars.  The proximity of the LMC (50 kpc)
permits individual massive stars to be observed under natural seeing
conditions (Evans et al. 2011), aside from R136, the dense star
cluster at its core, which necessitates use of adaptive optics or {\em
  HST} (Khorrami et al.  2017, Crowther et al. 2016). 
R136 has received particular interest since it hosts very massive stars 
($\geq 100 M_{\odot}$, Crowther et al. 2016), which are potential 
progenitors of pair-instability supernovae and/or merging black holes whose gravitational 
wave signature have recently been discovered with LIGO.

Star formation in the Tarantula began at least 15--30 Myr ago, as
witnessed by the cluster Hodge~301 whose stellar content is dominated
by red supergiants, with an upturn in star-formation rate within the
last 5--10 Myr, peaking a couple of Myr ago in NGC~2070, the central
ionized region that hosts R136. Star formation is still ongoing, as
witnessed by the presence of massive young stellar objects and clumps
of molecular gas observed with ALMA (Indebetouw et al. 2013). The
interplay between massive stars and the interstellar medium also
permits the investigation of stellar feedback at both high spatial and
spectral resolution (e.g.  Pellegrini et al. 2010).

\vspace{0.5cm}

\noindent {\bf MUSE observations of NGC~2070}

\vspace*{0.5cm}

\begin{figure}[htbp!]
\includegraphics[width=\columnwidth]{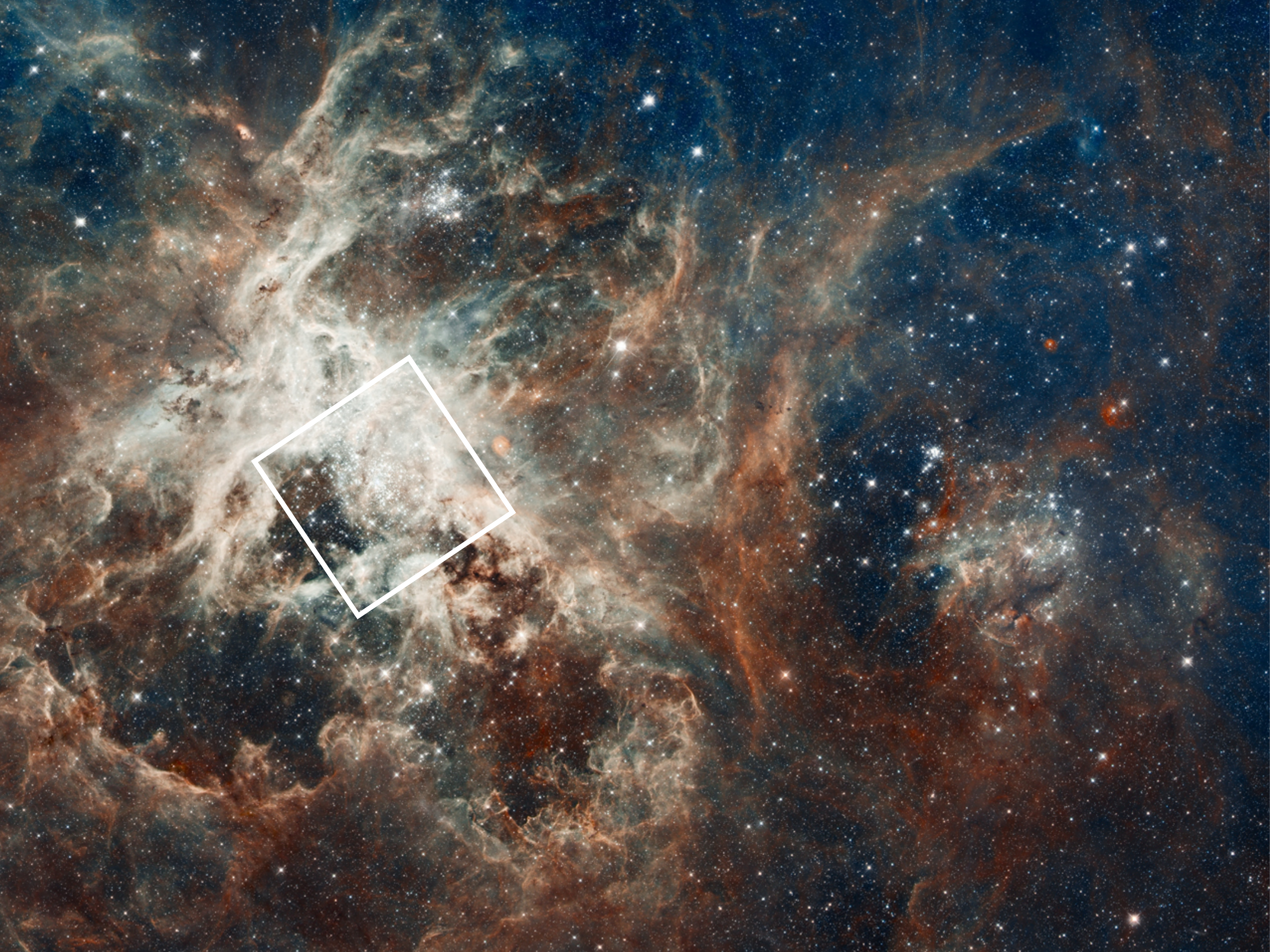}
\caption{MUSE 2$'\times2'$ mosaic (white square) superimposed on a 
colour-composite image of the Tarantula nebula (corresponding to $\sim 200 
\times 160$ pc) obtained with ACS and WFC3 instruments aboard Hubble. 
Image credit: NASA, ESA, D. Lennon et al. 
http://hubblesite.org/images/news/release/2012-01} 
\label{fig1}
\end{figure}

NGC~2070, the central region of the Tarantula nebula, was observed
with the Multi Unit Spectroscopic Explorer (MUSE) as part of its
original Science Verification programme at the VLT in August 2014.
MUSE is a wide-field, integral-field spectrograph, providing
intermediate-resolution ($R$\,$\sim$3000 at H$\alpha$) spectroscopy
from $\lambda\lambda$4600-9350 over 1$'\times1'$ with a pixel scale of
0.2$''$. Four overlapping MUSE pointings provided a 2$'\times2'$
mosaic which encompasses the R136 star cluster and R140 (an aggregate
of WR stars to the north) as shown in Fig.~\ref{fig1} on a
colour-composite of the central 200$\times$160\,pc of the Tarantula
obtained with ACS/WFC3 aboard HST.  The resulting image quality
spanned 0.7 to 1.1$''$, corresponding to spatial resolution of
0.22$\pm$0.04\,pc, providing satisfactory extraction of sources aside
from within R136. Exposures of 4$\times$600s for each pointing
provided a yellow continuum signal-to-noise (S/N) $\geq$50 for 600
sources, although a total of 2255 sources were extracted using
SExtractor; shorter 10s and 60s exposures avoided saturation of
strong nebular lines. Absolute flux calibration was achieved using
$V$-band photometry from Selman et al. (1999). An overview of the
dataset, together with stellar and nebular kinematics is provided by
Castro et al. (submitted).

\vspace{0.5cm}

\noindent {\bf Spatially-resolved nebular properties}

\vspace*{0.5cm}

\begin{figure}[htbp!]
\includegraphics[width=\columnwidth]{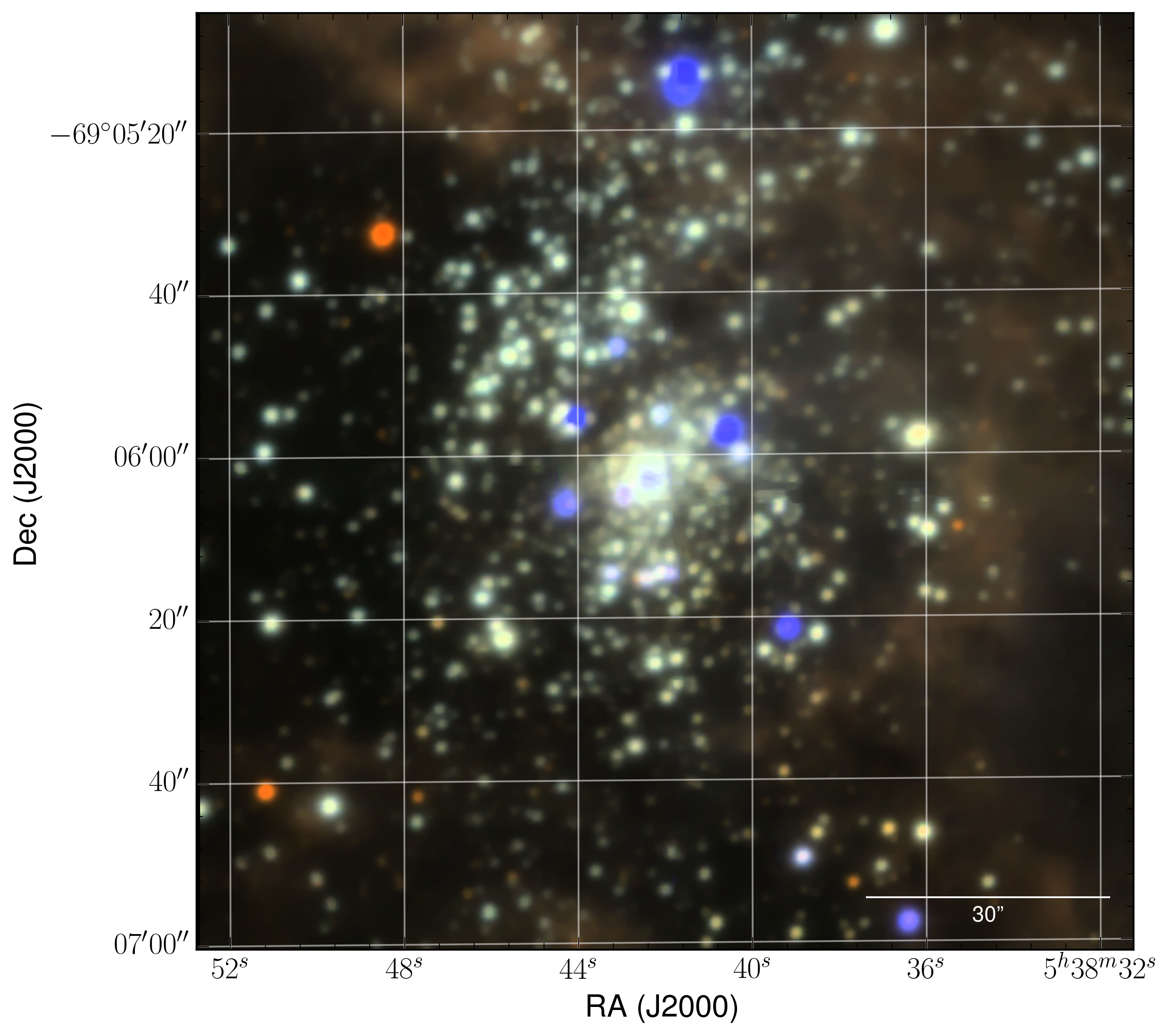}
\caption{{\bf (a)} VLT/MUSE colour composite of NGC 2070 (2$\times$2 
arcmin)
sampling $\lambda$6640 (red), $\lambda$5710 (green), $\lambda$4690 (blue).
Blue sources are Wolf-Rayet stars with prominent He\,{\sc ii} 
$\lambda$4686 emission, while orange sources are predominantly red 
supergiants.} \label{fig2}
\end{figure}

We present colour composite images extracted from the MUSE datacubes
in Fig.~\ref{fig2}(a) and Fig.~\ref{fig2}(b), highlighting the stellar
content and ionized gas, respectively. Fig.~\ref{fig2}(a) samples
$\lambda$6640, $\lambda$5710 and $\lambda$4690, such that most stars
appear white aside from cool supergiants (orange), such as Melnick 9
in the upper left, and WR stars which appear blue owing to strong
He\,{\sc ii} $\lambda$4686 emission, including R134 to the right of
the central R136 star cluster and the R140 complex at the top, which
hosts WN and WC stars. In contrast, Fig.~\ref{fig2}(b) highlights the
distribution of low-ionization gas ([S\,{\sc ii}] $\lambda$6717, red),
high-ionization gas ([O\,{\sc iii}] $\lambda$5007, blue) and hydrogen 
(H$\alpha$,
green). Green point sources generally arise from broad H$\alpha$
emission from WR stars and related objects.


Owing to the presence of ionized gas throughout NGC~2070, our MUSE
datasets enable the determination of nebular properties. Adopting a
standard Milky Way extinction law, there is a wide variation in
extinction throughout the region, with coefficients spanning
0.15 $\leq$ c(H$\beta) \leq$ 1.2.
On average, c(H$\beta$)=0.55\,mag, in excellent agreement with long-slit results
from Pellegrini et al. (2010).  Nebular lines also permit the
determination of electron densities and temperatures from [S\,{\sc
  ii}] and [S\,{\sc iii}] diagnostics, as illustrated in Fig.~\ref{fig3}. 
The dust
properties towards the Tarantula nebula are known to be non-standard,
with an average c(H$\beta$)=0.6 obtained from the
Ma{\'{\i}}z-Apell{\'a}niz et al. (2014) law and R = 4.4, although this
has little bearing on the nebular conditions determined here owing to the
use of red spectral diagnostics.

\vspace{0.5cm}

\noindent {\bf Massive stars in NGC~2070}

\vspace*{0.5cm}

MUSE permits the first complete spectroscopic census of massive stars within
NGC~2070, since previous surveys have been restricted to
multi-object spectroscopy using slitlets or fibres (Bosch et al. 1999,
Evans et al. 2011).  Spectral lines in the blue are usually employed
in classification of OB stars, so the $\lambda$4600 blue limit to MUSE
has required the development of green/yellow diagnostics.
Representative OB spectra from MUSE are presented in Fig.~\ref{fig4},
with classifications from blue VLT/FLAMES spectroscopy (Walborn et al. 2014).
Spectroscopic analysis of 270 sources with He\,{\sc ii}
$\lambda$5412 absorption is now underway using the non-LTE atmospheric
code FASTWIND (Puls et al. 2005), yielding temperatures and
luminosities from He\,{\sc i} $\lambda$4921 and He\,{\sc ii}
$\lambda$5412; preliminary fits to the illustrative spectra are also
shown in Fig.~\ref{fig4}.

Ultimately we will determine the properties of all the massive stars
in NGC~2070 to fully characterize its recent star-formation history,
substituting results from long-slit HST/STIS spectroscopy for the
central parsec of R136 (Crowther et al.  2016). Quantitative analysis
of the MUSE data should also provide useful insights into stellar
evolution theory. For instance, Castro et al. (2014) suggested
empirical boundaries for the zero- and terminal-age main sequence from
analysis of a large sample of OB stars.  The MUSE data will
enable a homogeneous analyis of a larger stellar sample, spanning a
broad range of evolutionary stages (i.e. main sequence, blue and red
supergiants, WR stars).

Of course, it is well known that most massive stars prefer company,
so it is likely that many of the MUSE point sources are multiple.
Fortunately, the majority of massive stars in NGC~2070 have previously
been monitored spectroscopically with VLT/FLAMES, revealing many
short-period systems. In addition, 30~Doradus also been the target of
a {\em Chandra} ACIS-I X-ray Visionary Programme (T-ReX), which has
monitored X-ray emission from the Tarantula over 630 days, permitting
longer-period systems to be identified. For example, Melnick
34, the blue (emission-line) star to the left of R136 in
Fig.~\ref{fig2}(a), has been revealed as an eccentric colliding wind
binary from T-ReX variability (Pollock et al. 2018).

\vspace{0.5cm}

\noindent {\bf Integrated spectrum of NGC~2070}

\vspace*{0.5cm}

In addition to spectra of the spatially-resolved stars and gas in
NGC~2070, it is possible to sum up the MUSE observations to provide
the integrated spectrum of the region. NGC~2070 would subtend 0.6$''$
if it were located at a distance of 10 Mpc, so MUSE offers the unique
opportunity to study both the spatially-resolved properties of an
intensively star-forming region and its cumulative characteristics.
The integrated spectrum of NGC~2070 is presented in Fig.~\ref{fig5}.
In addition to strong nebular lines, the high throughput of MUSE and
proximity of NGC~2070 reveals a plethora of weaker features in the
integrated spectrum, including the non-standard density diagnostic
Cl\,{\sc iii} $\lambda$5517/5537. Fig.~\ref{fig5} also highlights
broad blue (He\,{\sc ii} $\lambda$4686)
and yellow (C\,{\sc iv} $\lambda\lambda$5801-12) Wolf-Rayet (WR)
features in the integrated spectrum, with no evidence for a nebular
contribution to the former. These are often observed in the
integrated light of extragalactic star-forming regions.


Fig~.\ref{fig6}(a) compares the
strong line nebular characteristics of NGC~2070 with SDSS star-forming
galaxies and indicates similar high-excitation properties to Green Pea
galaxies (Micheva et al. 2017).
Analysis of the integrated spectrum reveals c(H$\beta$)=0.57 for a
standard extinction law, such that the de-reddened H$\alpha$ luminosity
is 1.5$\times 10^{39}$ erg\,s$^{-1}$, corresponding to 1/8 of the entire
Tarantula nebula (Doran et al. 2013). The current star formation rate 
(SFR) for NGC~2070 is 0.008 $M_{\odot}$\,yr$^{-1}$ adopting a standard 
Kennicutt relation between H$\alpha$ luminosity and SFR, modified for a
Kroupa IMF (division by a factor of 1.5), inferring a high star-formation 
surface density of  $\Sigma_{\rm SFR}$ $\sim$ 10 
$M_{\odot}$\,yr$^{-1}$\,kpc$^{-2}$. 
Conditions are  similar to clumps of intensively star-forming galaxies at 
high redshift, as demonstrated in Fig.~\ref{fig6}(b) which is adapted from 
Johnson et al. (2017).

\vspace{0.5cm}

\noindent {\bf Properties inferred from integrated light of NGC~2070}

\vspace*{0.5cm}

The inferred age of the region from the equivalent width of H$\alpha$
is $\sim$4 Myr, inferring a mass of 10$^{5} M_{\odot}$ for an
instantaneous burst of star formation, double the mass estimated for
the central R136 cluster. In reality, there is an age spread of 0--10
Myr for massive stars within the entire Tarantula nebula (Schneider et
al. 2018), although the peak of star formation is inferred
$\sim$4.5 Myr ago, excluding R136 with an age of $\sim$1.5 Myr (Crowther et
al. 2016). The H$\alpha$-derived ionizing output is 10$^{51}$ photon\,s$^{-1}$
for NGC~2070, corresponding to an equivalent number of $\sim$100 O7\,V stars.
This corresponds to $\sim$300 O stars for the nebular derived age (Schaerer \& Vacca 1998),
in good agreement with the number of MUSE sources displaying 
He\,{\sc ii} $\lambda$5412 absorption, albeit neglecting the (significant) 
contribution of the WR stars to the cumulative ionizing output.

We derive $\log$(O/H)+12 = 8.25 for NGC~2070,
adopting N$^{+}$ and S$^{2+}$ temperatures for singly ionized and doubly ionized
oxygen, respectively (the blue MUSE cutoff excludes the use of the 
stronger [O\,{\sc iii}] $\lambda$4363 line). Direct determinations for the entire 30 Doradus region indicate a somewhat higher oxygen
content (e.g. $\log$(O/H)+12=8.33, Tsamis et al. 2003). WR line luminosities are
metallicity dependent (Crowther \& Hadfield 2006), so adopting the LMC templates, one
would infer 20 mid WN stars and 5 early WC stars in NGC~2070, or N(WR)/N(O)$\geq$0.08.
This is in reasonable agreement with the resolved WR content of the MUSE field, namely
10 WN stars, 6 Of/WN stars and 2 WC stars. The rich star cluster R136 hosts four of the
most massive WN5h stars in the region, but only contributes  one third to the cumulative He\,{\sc 
  ii} $\lambda$4686 emission. In contrast the less prominent R140 complex -- host to two WN6 stars
and one WC star -- contributes  another  third of the He\,{\sc ii}  $\lambda$4686 emission and
dominates the integrated C\,{\sc iv} $\lambda$5808 and  C\,{\sc  iii} $\lambda$4650 flux.
This arises from the relatively weak wind strengths of main sequence WN5h stars, versus the 
significantly stronger emission from classical WN stars.

Strong-line calibrations are widely employed to infer the metallicity of extragalactic H\,{\sc ii}
regions owing to the faintness of auroral lines. Application of the commonly used calibrations
from Pettini \& Pagel (2004) would imply a SMC-like oxygen content of $\log$(O/H)+12 = 8.0
from both the N2 and O3N2 diagnostics, significantly lower than our direct determination.
If one had to rely on strong-line diagnostics for NGC~2070,
use of SMC-metallicity Wolf-Rayet templates from Crowther \& Hadfield (2006) would suggest an
unrealistically high number of mid WN stars, and, in turn, N(WR)/N(O)$\geq$0.3. This would represent a
severe challenge to current single/binary population synthesis models for a starburst region with 0.2 $Z_{\odot}$,
in stark contrast with N(WR)/N(O)$\sim$0.07 and 0.4 $Z_{\odot}$ obtained from our spatially resolved spectroscopy of the region.





\vspace{0.5cm}

\noindent{\bf References}

\vspace{0.5cm}

\noindent
Bosch, G. et al. 1999, A\&AS 137, 21\\
Castro, N. et al. 2014, A\&A, 570, 13\\
Crowther, P.A. et al. 2016, MNRAS 458, 624\\
Doran, E. et al. 2013, A\&A 558, A134\\
Evans, C.J. et al. 2011, A\&A 530, A108\\
Khorrami, Z. et al. 2017, A\&A 602, A56\\
Indebetouw, R. et al. 2013, ApJ 774, 73\\
Johnson, T.L. et al. 2017, ApJ 843, L21\\
Ma{\'{\i}}z-Apell{\'a}niz, J. et al. 2014, A\&A 564, A63\\
Micheva, G. et al. 2017, ApJ 845, 165\\
Pellegrini, E et al. 2011 ApJ 738, 34\\
Pettini, M., Pagel, B.E.J. 2004, MNRAS 348, L59\\
Pollock, A. et al. 2018, MNRAS 474, 3228\\
Puls, J. et al. 2005, 435, 669 \\
Schneider, F.R.N. et al. 2018, Sci, in press \\
Schaerer, D., Vacca, W.D. 1998, ApJ 497, 618\\
Selman, F. et al. 1999, A\&A 341, 98\\
Tsamis Y. et al. 2003, MNRAS 338, 687\\
Walborn, N. et al. 2014, A\&A 564, A40

\clearpage

\addtocounter{figure}{-1}

\begin{figure}[htbp!]
\includegraphics[width=\columnwidth]{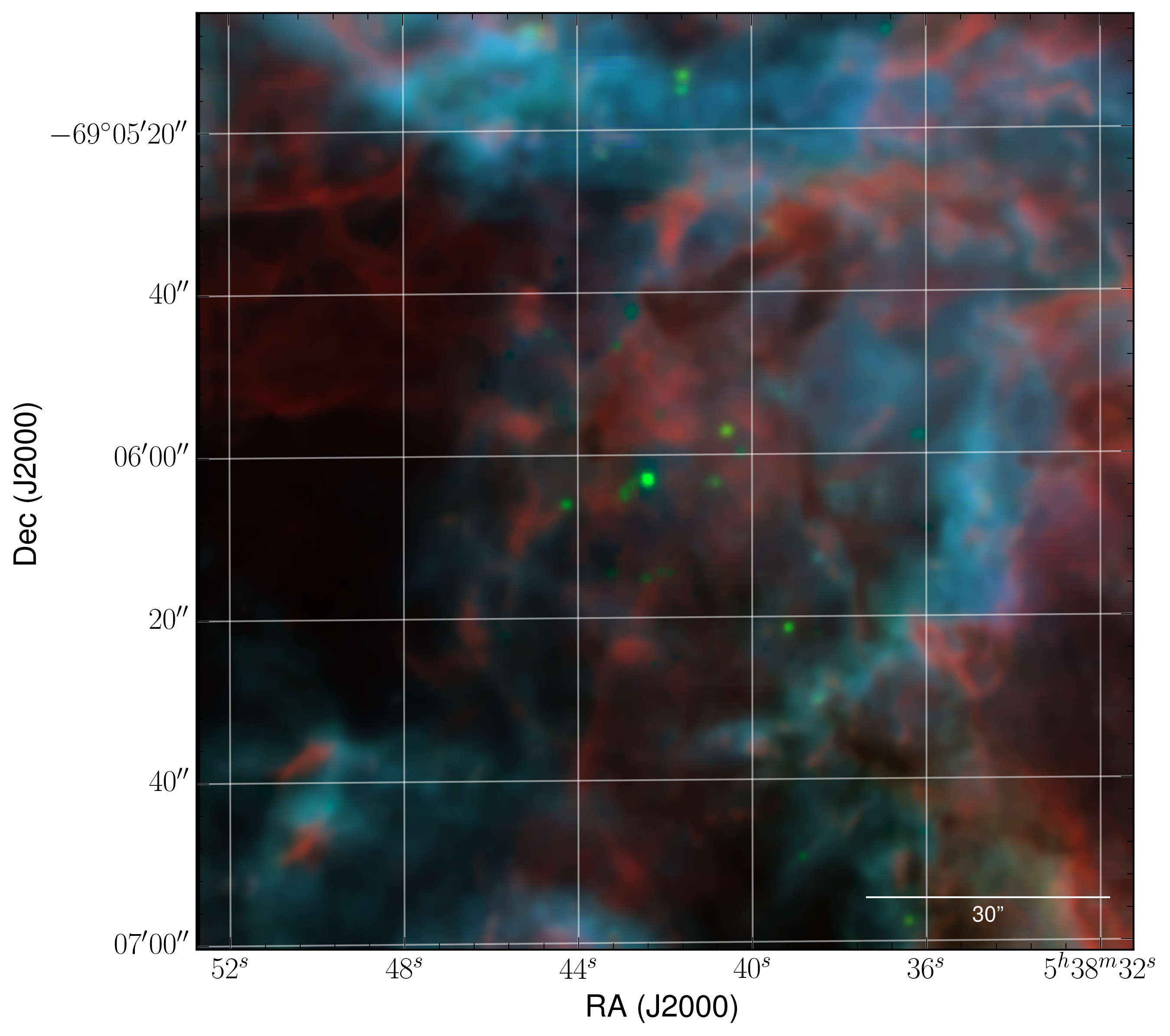}
\caption{{\bf (b)} VLT/MUSE colour composite of NGC 2070 (2$\times$2 
arcmin)
sampling [S\,{\sc ii}] $\lambda$6717 (red), H$\alpha$ (green), [O\,{\sc 
iii}] $\lambda$5007 (blue).}
\end{figure}

\begin{figure}[htbp!]
\includegraphics[width=\columnwidth]{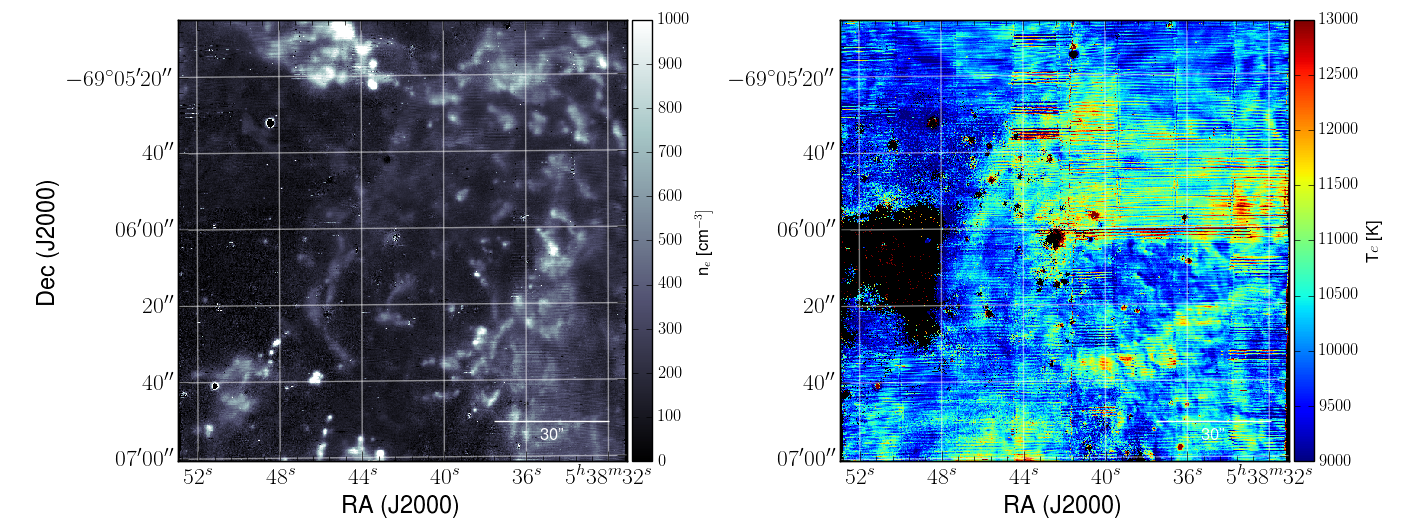}
\caption{Distribution of gas density and temperature within the MUSE
field of view, based on [S\,{\sc ii}] $\lambda$6717/31 and [S\,{\sc iii}]
$\lambda$6312/9069 diagnostics}
\label{fig3}
\end{figure}

\begin{figure}[htbp!]
\includegraphics[angle=-90,width=0.65\columnwidth]{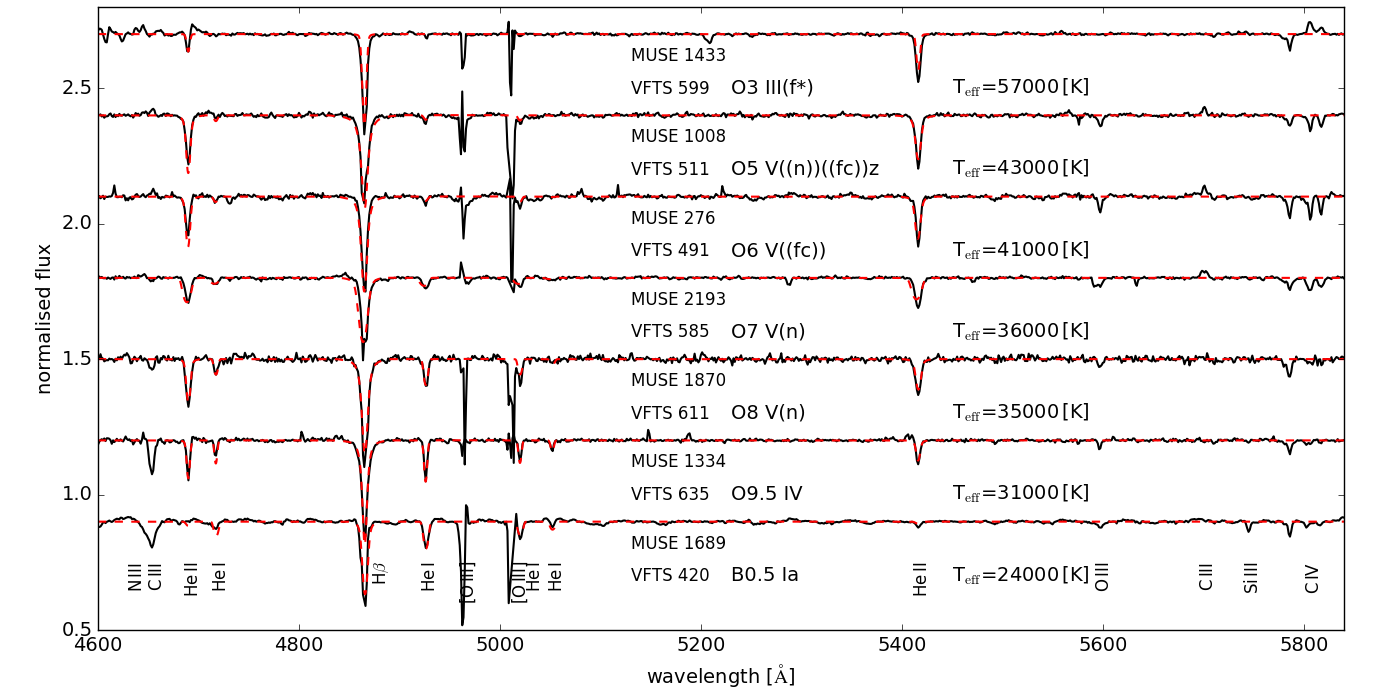}
\caption{Blue to yellow spectroscopy of representative OB stars in
  NGC~2070 observed with VLT/MUSE (black solid lines), including VFTS
  spectral types, and temperatures from FASTWIND model
    fits (dashed red lines) to He\,{\sc i} $\lambda$4921 and He\,{\sc
      ii} $\lambda$5412.}
\label{fig4}
\end{figure}

\begin{figure}[htbp!]
\includegraphics[angle=-90,width=0.6\columnwidth]{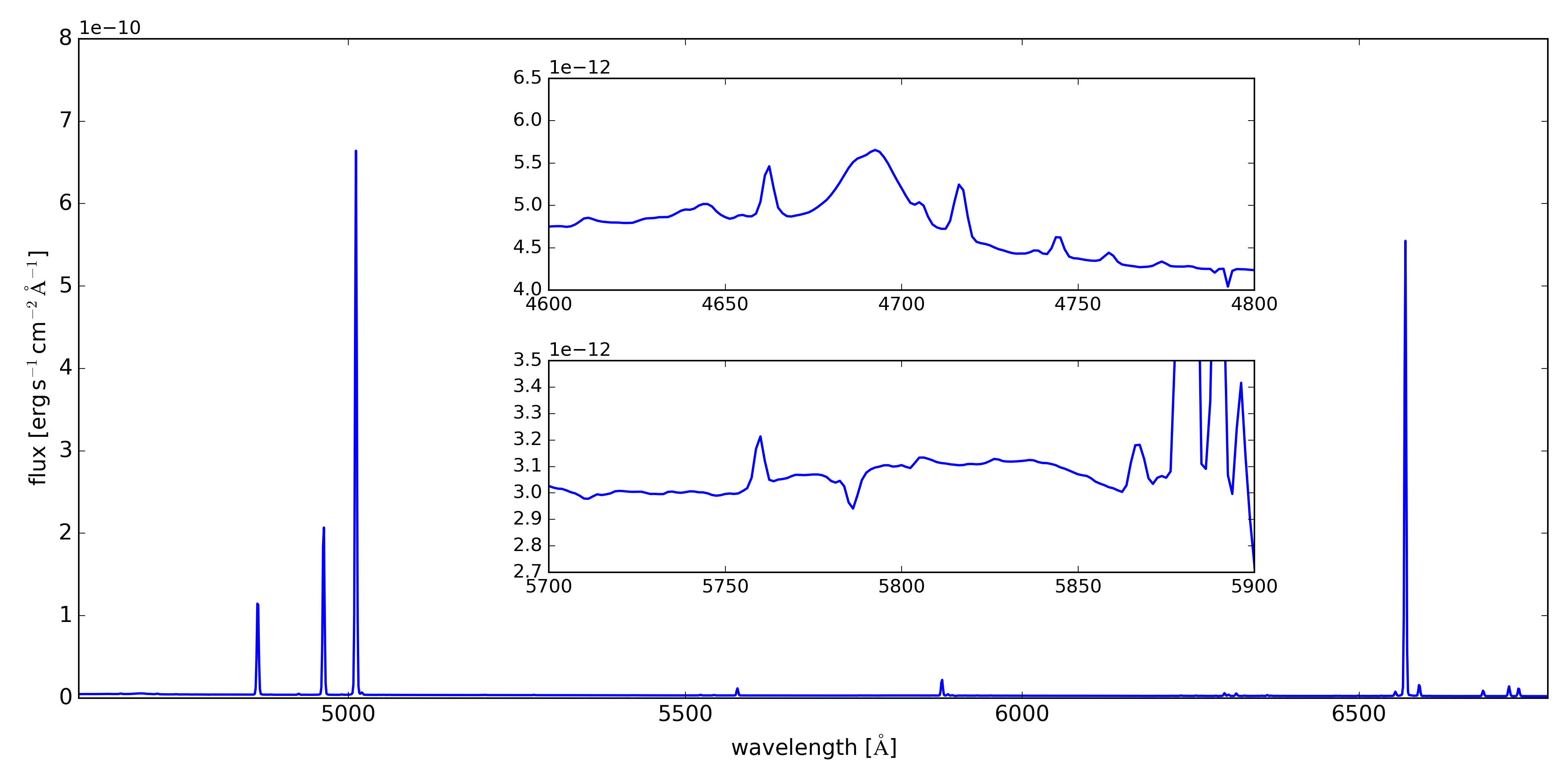}
\caption{Integrated MUSE spectrum of NGC~2070, revealing a striking emission line spectrum, with
characteristics reminiscent of Green Pea galaxies,  plus WR bumps in the blue (upper inset, He\,{\sc ii} 
$\lambda$4686 arising from WN stars) and yellow (lower inset, C\,{\sc iv} $\lambda\lambda$5801-12 owing to WC stars).} 
\label{fig5}
\end{figure}

\begin{figure}[htbp!]
\includegraphics[width=0.9\columnwidth]{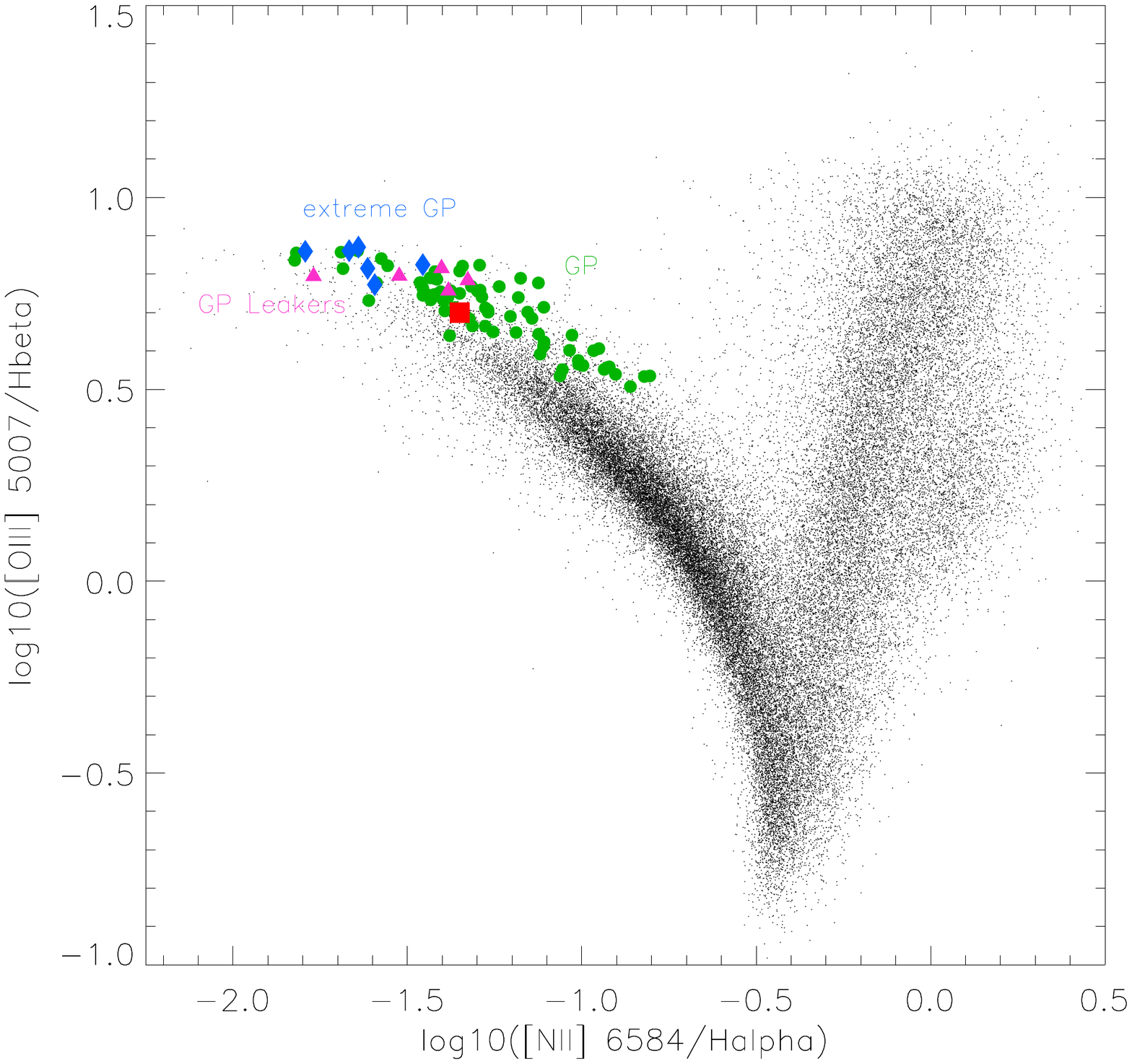}
\caption{{\bf (a)} BPT diagram illustrating the similarity in integrated strengths between 
NGC~2070/Tarantula (filled/open red square), Green Pea (green circles), 
extreme Green Pea (blue diamonds), and Lyman-continuum emitting Green Pea (pink triangles) galaxies, 
updated from fig.2 of Micheva et al. (2017), plus SDSS star-forming galaxies (black dots).}
\label{fig6}
\end{figure}

\addtocounter{figure}{-1}


\begin{figure}[htbp!]
\includegraphics[width=0.9\columnwidth]{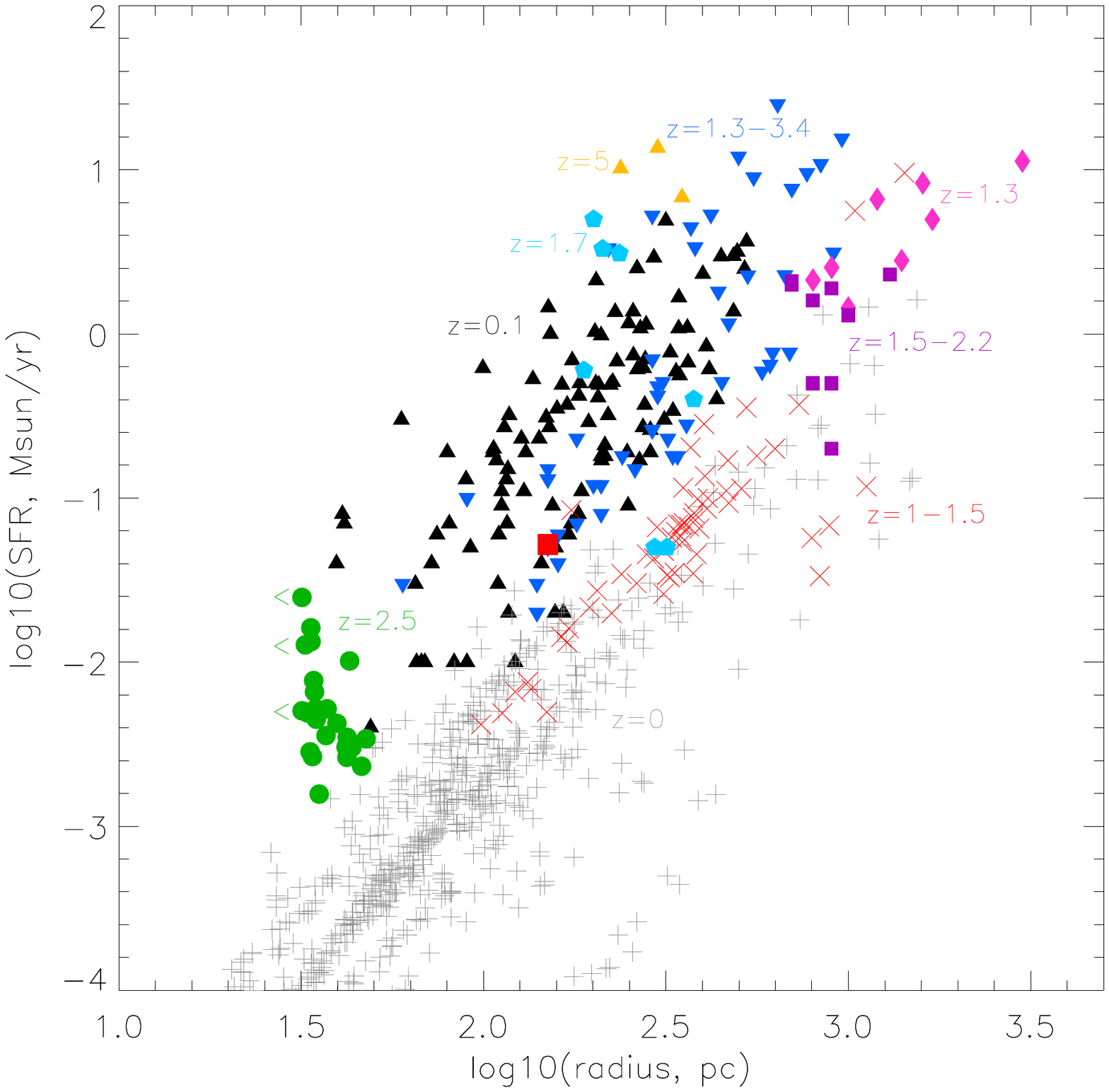}
\caption{{\bf (b)} Comparison between the integrated star-formation rate of NGC~2070/Tarantula 
(filled/open red square) and star-forming knots from galaxies spanning a range of redshifts, adapted from 
fig.2 of Johnson et al. (2017)}
\end{figure}

\end{document}